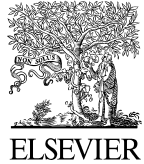

International Conference on Knowledge Based and Intelligent Information and Engineering Systems, KES2017, 6-8 September 2017, Marseille, France

# FindingPlaces: HCI Platform for Public Participation in Refugees' Accommodation Process

Ariel Noyman[a]*, Tobias Holtz[b], Johannes Kröger[c], Jörg Rainer Noennig[b], Kent Larson[a]

[a]*Massachusetts Institute of Technology, Media Lab, Changing Places Group, Cambridge MA*
[b]*HafenCity University Hamburg, CityScienceLab*
[c]*HafenCity University Hamburg, Lab for Geoinformatics & Geovisualization*

## Abstract

This paper describes the conception, development and deployment of a novel HCI system for public participation and decision-making. This system was applied for the process of allocating refugee accommodation in the City of Hamburg within the FindingPlaces project (FP) in 2016. The CityScope (CS) – a rapid prototyping platform for urban planning and decision-making – offered a technical solution which was complemented by a workshop process to facilitate effective interaction of multiple participants and stakeholder groups. This paper presents the origins of CS and the evolution of the tangible user interface approach to urban planning and public participation. It further outlines technical features of the system, including custom hardware and software in use, utilization in real-time as well as technical constraints and limitations. Special focus is on the adaptation of the CS technology to the specific demands of Hamburg´s FP project, whose procedures, processes, and results are reflected. The final section analyzes success factors as well as shortcomings of the approach, and indicates further R&D as well as application scenarios for the CS.

The authors would like to thank the many who participated in or helped with the development, design and deployment of FP. CS was firstly introduced at HCU CSL through two academic workshops lead by: Dr. Luis Alonso, Juanita Davis, Waleed Gowharj, Katja Schechtner, Katrin Hovy, Nina Hälker, Frank Rogge, Dr. Mario Siller, Dr. Agnis Stibe, Ira Winder, Ryan Zhang, Prof. Gesa Ziemer and the authors above.

* Corresponding author. Tel.: +001 617.253.5960
*E-mail address:* noyman@media.mit.edu

## 1. Introduction: The Need for FindingPlaces

### 1.1. Refugees challenge in Europe, Germany and Hamburg

In 2015, approximately 21 million people were fleeing from their home countries. A total number of 1.2 million applications for asylum were filed in Europe, an increase of more than 300% to the previous year [1]. As the European Union struggled to determine a distribution key on which basis the member states would receive refugees, enormous disparities emerged within the EU: Germany received more than a third of the asylum applications (442,000), more than any other member state and more than ever registered by the German Federal Office for Migration and Refugees [2]. The persistent influx of asylum seekers posed major challenges for German federal states and municipalities. As a consequence, available solutions were ad-hoc implemented and in many cases refugees were accommodated in tents, warehouses or gymnasiums.

In addition to this time-pressure, the limited space for refugee accommodation represented a serious obstacle in a long-term planning process, especially in densely built cities and rapidly growing urban regions. Although a public demand, many cities have failed to guarantee fair and equal distribution of refugees within their bounds. In Hamburg, accommodation facilities concentrated in certain neighborhoods while others received little to no refugees at all. Civil protest against refugee accommodation projects began to arise. Only few cases of physical hostility towards foreigners were reported but nevertheless highlighted the civics' demand to be heard in regards to refugees accommodation. The heated and emotional debate motivated Hamburg's First mayor to initiate a citizen dialogue to openly discuss where and how to accommodate refugees in the future. The idea was to treat the issue as collective and city-wide challenge, in which citizens themselves could take responsibility and contribute to a common solution. This participation process was named FindingPlaces (FP).

### 1.2. Project Mandate for the HCU-MIT City Science Collaboration

In June 2015, Hamburg's First Mayor Olaf Scholz and MIT Media Lab Director Joi Ito signed a long-term research collaboration agreement that promoted the establishment of the City Science Lab (CSL) at HafenCity University Hamburg (HCU) [3]. CSL was conceived as a Living Lab – an autonomous and self-sustaining extension of the research conducted by MIT's Changing Places Group (CPG). CSL inherits MIT Media Lab's 'consortia' structure, in which industry leaders and private companies come together to promote research in mutual fields of interest. At MIT, more than 80 different consortium members support the Media Lab and enjoy privileged access to research and intellectual property [4].

By February 2016, the HCU and MIT collaboration was to take a new path as mayor Olaf Scholz assigned CSL the development of a participation process that would enable citizens to engage in finding accommodations for a predicted influx of ~79,000 refugees in the city [5]. The goal was to incorporate the citizens' personal experience and local knowledge into the political and administrative evaluation of potential locations. The results and proposals emerging from the participation process were to become recommendations for political decision-making. Thus, FP was developed in close coordination with the Senate Office, the Central Refugees Coordination Staff (ZKF), district administration representatives and the Hamburg Urban Development and Revitalization Agency (steg), a company specialized in citizen participation processes who moderated the public engagement. The mayor's office allowed three months for conception and development of FP.

### 1.3. Project Idea

The FP team envisioned a design of highly trans-disciplinary character. The need for involving Hamburg´s citizenship and utilize their valuable local knowledge asked for a new participatory decision-making process. In addition, complex urban planning data needed to be processed to enable an effective search for suitable land-parcels. Prior to FP, reliable decision-making for the allocation of refugee accommodation was done by experts based on broad technical, legal, and contextual knowledge. To enable citizens' input in such complex tasks, MIT´s CityScope (CS) was proposed as a decision-making and knowledge-support tool. Featuring an advanced Tangible User Interface (TUI), CS is able to present relevant information in an easy-to-comprehend and easy-to-interact manner. Merging these concepts, a series of public participation workshops was planned that centered around interactive CS stations displaying task-related data to citizen groups as they worked out decisions. The main conceptual components of FP included 1) a workflow design for the overall workshop series, 2) a choreography („procedure") for the individual participatory workshops, 3) the technical adaptation of CS interactive tables, and 4) extensive pre-processing of urban data.

## 2. CityScope: An Urban Simulation Platform

### 2.1. CityScope Background

CityScope (CS) is an ongoing research theme taking place at the MIT Media Lab's Changing Places Group (CPG) in the past several years. CS builds upon a long-term research effort dating back to 1999 with the 'Augmented urban planning workbench' and the 'Illuminating clay' table [6] by Underkoffler, Ishii & Ben-Joseph as well as Larson's 'Louis I. Kahn - Unbuilt Ruins' exhibition [7] which utilized location tagged-building objects; these projects shared a common goal to demystify spatial design and analysis using tangible-computational platforms.

CS features many different iterations of an urban simulation platform with the main goal of making complex urban questions accessible and tangible to various audiences. This goal differentiates CS from other highly-specialized, experts-focused planning tools: Its tangible, user-oriented design prompts a discussion not limited by expertise or prior knowledge [8]. CS development aims to increase possible complexity and diversity of the urban questions handled by a single CS platform while improving user experience, interface and interaction through the elimination of unnecessary complexities. CS developers aim to excel these two efforts in concert to provide balanced complexity and legibility [9]. CS projects are either demos (experiments and research done at MIT Media Lab, usually in the form of generic and scalable tools) or deployments (CS platforms that serve as active tools in local planning processes around the world). In the past few years, deployments took place in the UAE, Australia, China, Andorra, Boston, Taiwan as well as in Hamburg's FP project.

### 2.2. Generic CityScope Set Up: Technical Description

A common CS platform would feature a few key components: A tangible urban model (city, neighborhood or street scale) poised over a table frame, a computational analysis unit and a feedback module. A CS table usually includes a set of color-tagged bricks acting as intractable buildings or massing elements [10]. The computational analysis unit has sensors or cameras and computers for real-time scanning of the scene. The feedback module contains display screens, projectors and occasionally other representation tools (AR, VR or touch feedback). As in the case of FP, certain deployments require a more complex setup, with multiple components and feedback modules.

### 2.3. CityScope as a Community Engagement Platform: Case Studies

Since its initial conception, CS went from a proof-of-concept to a functional and deployable urban intervention and decision-making support system. A couple of CS iterations cemented its relevancy as a viable public participation apparatus and encouraged CSL to build upon CS concepts for Hamburg's refugees challenge: (1) an observational study on the usability of CS in community-led planning process and (2) a public participation process for planning Bus Rapid Transit system in Boston.

CS PlayGround This project aimed to examine an alternative community planning process for the emerging redevelopment of Kendall Sq. and MIT east-campus. CPG team has developed a new, large scale CS platform that depicted several urban blocks at the far east side of MIT, and allowed users to virtually design this area with a responsive and dynamic zoning mechanism. During the platform's development, it became clear that an unbiased outlook on the performance and usability of such system is necessary in order to compare it to common planning practices. An observational study was preformed, aiming to examine its usability in a simulated urban planning process and to investigate its effect on stakeholders' engagement and decision-making. The observational study was conducted with a representative sample of users from different backgrounds. The users were invited to examine the usability of the collaborative tangible interface and to express their opinion on ways to better planning processes through such platforms. Participants from MIT community, including students, staff, and affiliates were invited. Upon arrival, the participants were divided into two groups: one group engaged in a 'pen & paper' discussion (common community meeting session) while the other was employing interactive tangible model of the urban area (CS session). In both sessions, participants were asked to respond to real-life planning and design challenges in respect to existing plans and regulations imposed on site. The sessions were video recorded and analyzed; a coding scheme was developed for analyzing the video observations to examine the wide spectrum of actions, verbal cues, and nonverbal gestures, and quantify the occurrences of these interactions during the experiment.

Researchers later analyzed the video recordings of both sessions using a predetermined coding scheme which included several types of actions, verbal and nonverbal behaviors, verbal communication and physical gestures. The analysis showed a clear favoring of the TUI approach over classic method. The study showed that, for example, dealing with physical objects made participants more active and engaged in the planning and decision-making process. Furthermore, due to the feedback visualization, participants could quickly share a common understanding of the proposed plan. As well, real-time design feedback (such as zoning violation or exceeding limits) helped participants in the sense-making and learning process. At last, this research suggested that TUIs such as CS are superior to traditional public participation for urban planning in terms of rapid prototyping and feedback, collaboration, and decision-making process [11] [12].

<u>CS Boston BRT</u> In early 2015, CPG collaborated with the Mobility Futures Collaborative of the MIT Department of Urban Studies and Planning (DUSP) to develop interactive tools aimed to communicate possible impacts of new BRT (Bus Rapid Transit) systems in Boston. The team tested these tools in a series of community engagement workshops, allowing the general public to interact, learn and co-create better transit solutions for their needs. The tools were designed to bring key stakeholders from the community, non-governmental organizations, government, and planners to engage in constructive discussion, and to encourage participation through technology. By developing an understanding of BRT concepts, trade-offs, and impacts, users could also use the tools to design their own proposals, and thus contribute their ideas to future planning. The tools were designed to depict impacts of BRT corridor implementation at three scales: the region, the neighborhood and the street. The region was represented through a web-based tool presented on a large vertical touchscreen that allowed users to explore transit system performance region-wide. The neighborhood and street scales were represented with two CS platforms which examined how different corridor types might perform in a neighborhood, or how different street elements (such as bus station types or bike lanes in lieu of parking) might impact the streetscape and corridor performance.

Six facilitated workshops were carried during October 2015. Each workshop lasted approximately two hours and followed a similar agenda and approach, including registration, orientation, exploration using the tools and post-workshop survey. Sessions were held with the help of community facilitators who led the discussions at each tool station; Their aim was to reduce participants' potential perception that control of the technology, and therefore the analyses, is in the hands of the tool developers. An open, interactive and peer-led process intended to empower all participants to play an active role in the planning exercise.

Through exit surveys and post-workshop meetings, participants reported having learned 'a great deal' and provided evidence of deeper, longer-term learning potential. The evidence collected during the workshops show how the tools contribute to learning while enabling high levels of interaction and conversation with others or questioning of the tools and their assumptions. The tools and their implementation had strengths and weaknesses: Multiple scales of interaction enhanced learning and co-creation; The tangible models suffered from limited possibilities for interaction as only several pieces could be interacted with, constraining potential creative exploration; In none of the scales outcomes produced by the users were documented (e.g., preferred corridor type or alignment and willingness to trade-off bus priority for parking spaces). Several major lessons learned from the BRT project: The importance of using real-time, multi-scale representation in order to facilitate a multifaceted discussion; the need for clear, simplified and straightforward goals for facilitated workshops; and the importance of using well-trained and preferably local moderation teams. Many of these conclusions found their way into FP and allowed a more concise community engagement experience.

## 3. Adopting CityScope for FindingPlaces

### 3.1. Concept

   Adapting CS for the FP project required multiple alterations in several aspects; At the time, existing CS hardware and software were limited and could not fulfill the challenging scope of FP. Through the development phase, CS main concepts, setup and interactions were inherited and all software and hardware were replaced with tailored solutions to FP, including: networked communication between multiple devices, integration of a Geographic Information System (GIS) and persistent data management. Workshops were designed to focus on one interactive CS table showing a user-selected part of the city on a scale of 1: ~750m. Geographic data on Hamburg's smallest administrative units (parcels), were enriched by any available and relevant data, such as general ownership or land use to allow citizens to discuss these places' suitability for refugee accommodation in an informed manner. Using CS, users could place a specific data brick onto a projected map and query attributes of the underlying geospatial data. A selection of data bricks assigned to different pre-defined housing values could be placed to add proposed numbers of accommodation to a certain parcel. The results of these interactions were visualized on auxiliary displays in real time, using maps of different scale and granularity, diagrams and statistics. For example, if a user placed a data brick representing an accommodation capacity of 50 people, its location and density would be displayed on the table itself, on a map projected onto another table featuring the entire district and on a regional map of Hamburg projected on a wall-mounted canvas. In addition, multiple diagrams and statistics representing the totals of accommodation capacity for a specific district and for the entire city of Hamburg, would be displayed on nearby monitors.

### 3.2. Diagrammatic Structure and Technical Description

**Figure 1.** System Components and room setup

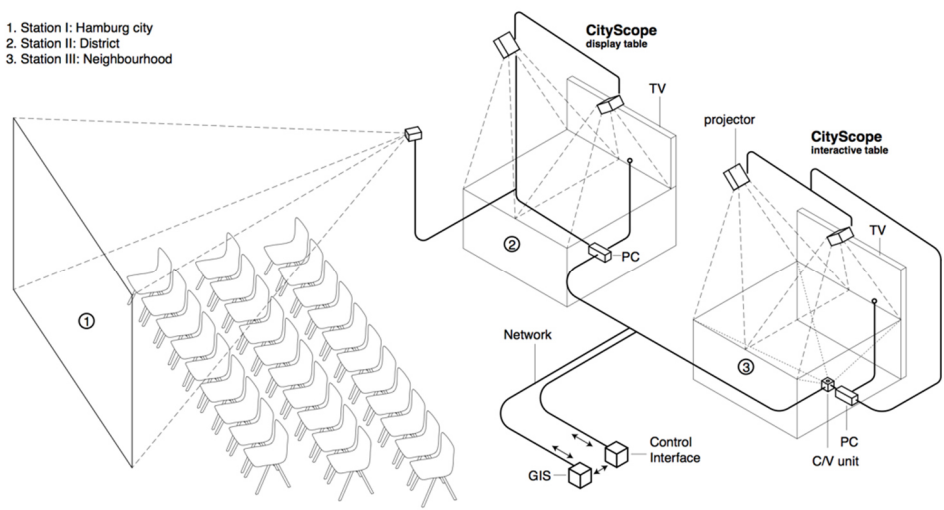

   The FP set up included three main components: Image processing of live video data to interpret the users' interactions with the CS table, translation of those interactions into a geospatial context (GIS), and communication and visualization of the effects of the interactions. Additional system management and operator controls had to be developed for interactively managing the system to suit each workshop. The tangible element of the system utilized the ease of recognizing and localizing few regularly shaped, color-coded objects on a fixed grid system: In FP, small objects were assembled from colored LEGO bricks so that users could pick and place on a transparent table surface. These objects were color-coded from bellow in regular, square patterns. The placement of these data bricks onto the table was constrained by fabricated square grid. All other grid cells were filled with low, neutrally white objects that provided a canvas for the map display via the overhead projectors. Cameras monitored the underside of the transparent table surface through which the color-coded bottoms of data bricks were visible. Shapes visible in the video feed were then analyzed and compared against a lookup table of known brick color-codes. The data attribute(s) and coordinates of successfully detected bricks were then transmitted to other components of the system. These analyzed the state for changes and translated it into the GIS context. If changes were detected, the maps and statistics were updated accordingly on all displays.

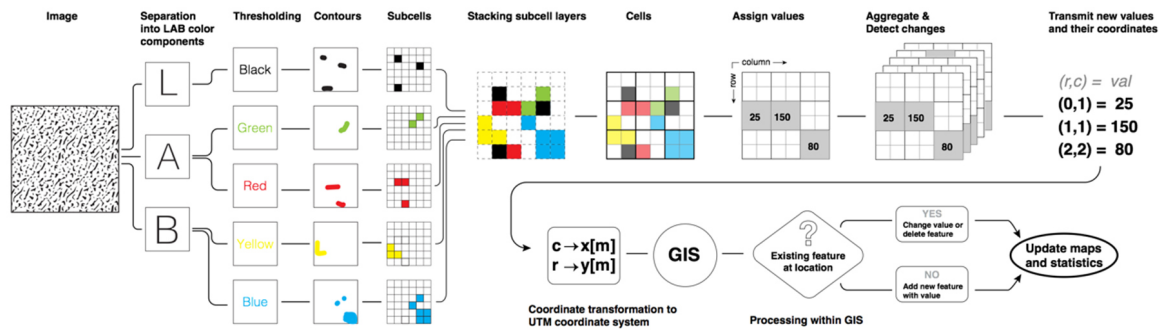

**Figure 2** Processing of a video frame from object detection to interpretation and handling in GIS

<u>FP Scene setup</u> While initially two tables were equipped with interactive CS components, only one ended up being used as such. The other table was used to depict a dynamic overview map of the specific district discussed during the workshop. Each square table was illuminated by two projectors, each covering one half. Auxiliary displays were handled by large TV screens with a wall-mounted canvas showing a global overview of FP statistics and information.

<u>Scanning & computation components</u> Each table was equipped with a workstation, driving the nearby displays and projectors. The workstation at the district table was used to perform most backend operations from image recognition to GIS server communication. The workstation at the interactive table was used to control the map displays. The GIS server was hosted on a virtual machine on the university's servers. Connectivity between the devices was provided by the HCU's LAN network. To be able to scan the entire bottom surface of the table simultaneously, four cameras were used, each one covering a square quarter of the table. Image recognition was handled by two single-board computers which streamed the video from the webcams to a workstation computer.

## 3.3. Software, Tools, Custom Code

Given the project's tight schedule, existing open-source solutions were employed to ease the burden of development. The software system was separated into a microservice-like structure so it could be developed in a modular way, allowing asynchronous rebuilding or replacing of components. The backend was developed in Python, the frontend with standard web technologies (JavaScript, HTML, CSS). The majority of communication between devices and processes was handled using a publish-subscribe pattern built on the Crossbar router with Autobahn|JS and Autobahn|Python clients using WebSockets, providing asynchronous processing. For example: one of the pub/sub topics was the current map extents while the other was the current global workshop statistics regarding refugees' accommodation numbers.

The main GIS server was an instance of GeoServer, with its data managed in a PostgreSQL/PostGIS database. Map display on the tables was realized via OpenLayers clients in full screen web browser windows. OpenLayers was chosen as highly capable and well-proven client library for browser-based interactive maps. Data transfer and spatial queries between the GIS server and the clients were performed via GeoServer's WFS API using the Python bindings of OGR/GDAL, while specific non-spatial queries were run directly against the database. Figure [2] shows the process of interpreting a video frame up to data handling in the GIS. The single-board computers ran ffserver and ffmpeg to stream their video feeds to the network. Video processing and interpretation were implemented using the computer vision library OpenCV and numpy.

To evaluate the general availability of a parcel and to enable users to form educated opinions about its suitability for refugee accommodation, detailed information about the parcels of Hamburg had to be available. Many data sources were aggregated into the GIS database, such as information about which parcels are city-owned and thus available for discussion, known current uses of parcels or existing constraints such as nature reserves. These geospatial data layers were intersected and weighted to create initial assessments of places, then sorted into three ranked classes:

- *High indication of unsuitability:* Places significantly affected by highly restrictive criteria such as nature conservation, cemeteries, significant noise emission, etc.
- *Medium indication of unsuitability:* Places insignificantly affected of aforementioned highly restrictive criteria but significantly affected by less restrictive criteria like parks, designated recreational areas, proximity to high-voltage power lines or similar.
- *Low or no indication of unsuitability:* Places not affected by highly restrictive criteria with less restrictive criteria affecting less than 50% of the area.

These classes were set to provide a reasonable baseline for the selection of places to discuss in detail during a workshop. Additional geospatial data hosted by external providers was integrated as cascaded WMS in GeoServer (e.g. aerial imagery of Hamburg) or client-side tile layers in OpenLayers (e.g. OpenStreetMap). These only served viewing purposes.

## 4. FindingPlaces Procedure

### 4.1. Overview

Between May and July 2016, a total of 34, two-hour workshops were held at HCU with nearly 400 participants. Each workshop focused on one of the city's seven districts. The workshops were advertised via various media channels and ~40,000 brochures distributed all over the city, having an overall reach of ~5 million citizens. Participants were asked to register online; up to 20 people per session were eventually invited, due to space limitation and CS platform dimensions. On average, 11 people participated per workshop. There were no restrictions to registration. However, participants could only attend one workshop per district in order to allow others to participate too. Interest, response and registration numbers varied depending on the district. The diverse range of participants was characterized by a strong heterogeneity concerning age, profession, political views, social values and personal motivation to participate.

### 4.2. Venue and Moderation

The workshops took place from Monday to Saturday at different hours in order to reach as many participants as possible. The venue – a gallery space of about 150m² on HCU´s entrance floor – was deliberately chosen for being a neutral location for the often emotionally charged discussions on refugee accommodations. A team of six to eight people was in charge of the organization and conduct of the workshops: one moderator leading the discussion, one assistant documenting the exchange, one CSL-researcher accompanying the workshops for scientific purposes, one technical staff operating the equipment and one or two representatives each from the Central Refugees Coordination Staff and district administration. The moderators were substantially leading the workshops, while other team members mostly provided on-demand details concerning CS, specific sites, or on the topic of refugee accommodation in general. The open and facilitated discussion contributed to the participants' grasp of the complex subject matter.

*4.3. Description of Engagement*

The venue of the workshops was divided into three stations, displaying the city on different scales: entire city (I), district (II) and neighborhood (III). (see figure [1])

Station I: Hamburg City First, an animated movie was shown to inform participants about the workshop procedure, the context conditions of the refugee influx, and the challenge of providing appropriate accommodation. A task was presented as a guiding question: Which city-owned parcels would be suitable locations for refugee accommodation? The overall ambition was to find sites that in total could accommodate 20,000 refugees and thus meet the predicted demand until the end of 2016.

The locations of all existing and planned refugee accommodations were shown on a map, as well as the current statistics on the refugee distribution to the different districts. In addition, the targeted number of 20,000 accommodation places was displayed, counting down in real-time in response to new locations proposed by the participants during the workshops.

Station II: District After the introduction, the first CS was introduced. Here, a satellite image projected onto the table showed the city district at stake. Street and neighborhood names as well as distinctive points of interest were displayed to provide orientation. As done before on city scale, the locations of existing and planned refugee accommodations including their actual occupation rates were indicated. The public spaces were colored according to their previously determined suitability classes: Red for places with a high indication of unsuitability, orange for a medium indication and yellow for likely suitable places. Further information on the number of inhabitants and refugees currently accommodated in the district was presented on a screen. Taking into account that not all spaces could possibly be discussed during one two-hour workshop, the participants were asked to select focus areas to zoom-in at the next station. Selection criteria could be specific knowledge of a place, the need for equal distribution of refugees within the city, or the color-index of the spaces.

Station III: Neighborhood The chosen focus areas were projected onto the second interactive CS, the suitability classes shown in colored hatching. Here, participants were able to identify details like buildings, parks or playgrounds. In addition to refugee accommodations, social infrastructure like kindergartens or schools as well as bus or train stops was shown. By placing a marker piece onto a parcel on the CS, detail information on the property was displayed on a screen (area in m², planning regulations, designation of the area, as well as restrictions such as nature conservation, biotope, or high-voltage lines). Additionally, the potential accommodation capacity was computed and indicated. Participants´ verbal discussion of the pros and cons for each parcel were logged and displayed on the screen.

If participants managed to identify a location suitable for refugee accommodation, this parcel could be suggested to the city officials. To do so, another marker piece was placed on the CS, indicating the number of accommodation places (between 40 to 1,500). At that point, the screens and the CS at the Hamburg and district-stations were also displaying the location of the suggested parcel and the proposed number of refugees was included in the statistics.

*4.4. Feedback and Commenting Loop*

After each workshop, a list of the suggested parcels was handed directly to the Central Refugees Coordination Staff along with all information regarding the planning restrictions as well as transcripts of the discussion. These materials were also added to the official FP website allowing public access. The Central Refugees Coordination Staff initiated a screening process in which each suggested parcel went through a feasibility test. The results of this process as well as their reasoning were published online within two weeks. Parcels which were deemed suitable were then further studied by the planning authorities.

In total, 161 locations were suggested by the participants and evaluated by the authorities. With these, accommodation solutions for almost 24,000 refugees were proposed, exceeding the initial targeted goal of 20,000. More than half of the parcels were designated parks, green areas in inner-city locations, landscape, or agricultural spaces in rural areas, that are mostly subject to nature or landscape conservation. Another 15% of the suggested parcels were used as sports fields or playgrounds. Others were parking lots, commercial and industrial areas, parcels designated for future housing projects or port area parcels. Almost three quarters of the suggested locations were rated as not suitable in the initial assessment, leaving 44 rated as feasible. A further 24 were excluded after a detailed examination. Ultimately, 6 received recommendations for implementation and 10 were taken into consideration for future planning. In these lasting 6 locations, approximately 750 refugees could be accommodated, ~4% of the initial target.

Several reasons led to rejection: More than a third of the parcels were not available due to other land-uses such as commercial activity or sport and leisure purposes. Another third could not be used due to direct conflicts of use (mostly parks and playgrounds). Other reasons for rejection were of technical or structural nature, e.g. hazardous environment, contamination, topographic constraints, protection of historical monuments or a lack of nearby public transport and social infrastructure.

*4.5. Interaction Process*

Since CS was used in a participation process of this kind for the first time, none of the workshop participants were familiar with the functionality of the platform. Some feared that the data used in the platform might have been manipulated to politically control the suggestions in the workshops. Hence the origins of the used data as well as the criteria that led to the classification of the space and planning restrictions had to be clearly explained.

The workshop methodology and technology enabled a high level of involvement and direct discussion between experts and non-experts leading to evidence-based and goal-oriented interaction. Despite the emotional charge and heated public debate regarding refugee accommodation, the discussions and atmosphere in the workshops were mostly calm and constructive. Occasionally, participants expressed their dissatisfaction with the refugee policy pursued by the Hamburg senate. Some were arguing that FP was 'faked' participation, only utilized to restore peace to the city. However, it could be observed that the disapproving comments were mostly based on vague or incorrect information and were hence quickly diminishing with the provision of more precise information and the discussion of specific locations in the process. This points out that the dialogue between the participants could be rationalized by shifting from a theoretical discussion to a more tangible level through clear visualization of facts. The workshop results owned a high level of practicality as concepts were based on information thoroughly prepared and presented. In that sense, participants could not reject a proposed parcel without providing arguments that were clear to other participants. During the workshops, many ideas of an ideal solution for refugees accommodation had to pass a 'reality check', pushing some to be revised as they were incompatible with the applicable planning regulations and restrictions.

## 5. Results Discussion

<u>Strengths</u> By both stakeholders and participants, FP was evaluated as a positive experience, and the CS tool was recognized a highly supportive instrument for public participation and real-time decision-making. FP succeeded especially on the 'soft' level of human interaction: Citizens felt as partners in an 'eye-level' dialogue with policy makers and city administration, being able to supply planning authorities with relevant information based on their local knowledge. The project built up acceptance towards refugee accommodation in Hamburg and triggered high-quality feedback. Making administrative procedures and decisions transparent effectively contributed to the 'political literacy' of the general citizenship.

<u>Weaknesses</u> A key challenge was the tight schedule in which the project needed to be implemented. Further, logistical limitations reduced the overall exposure of the CS tool. Due to the size of the tables, workshops were bound to be held at HCU; This naturally reduced the number of potential remote participants, thus contributing to a selection bias which is well-known in public participation projects.

Another constrain was the lack of available urban data. Despite thorough pre-processing of urban data, non-expert participants had trouble understanding the professional planning content. As participants were not used to working with maps and satellite images, orienting the projected images and assessing them adequately was difficult.

<u>Opportunities</u> FP demonstrated the capacity of CS as a decision-making tool both for citizen participation as well as expert planning. The tool can adopt to a range of problems similar in type yet different in context, e.g. in real estate development or social infrastructure analysis. CS unfolds special potential in common multi-stakeholder participation processes such as urban planning or development strategies. In all cases, CS can effectively support the discussion of multiple scenarios with playful and tangible interfaces. The enhancement of data resources and visualizations (for instance by projecting higher resolution images or by integrating street views) will yield new scenarios for practical application as well as for technical development.

<u>Threats</u> Lack of public recognition and opposition from professional stakeholders threatens CS research and innovation. In addition, insufficient data supply may render the tool useless. Methodological shortcomings (selection bias, lack of transparency about participants´ roles) obstruct the reliability and acceptance of the approach, too. Pre-processing of data allows for manipulation of the overall process outcome (e.g. suggestive color-coding for map visualizations influencing the direction of discussions). Public communication of project results without further information about scope, limits, and the workshop events may turn out misleading or invalid. If a project is of high public or political interest – as the case was with FP – the approach runs danger of becoming instrumentalized by political forces or interest groups.

## 6. Conclusion

Applying a CS platform for the FP refugee accommodation project has shown how digital technology can effectively support social challenges and physical changes. The shortcomings as highlighted in section 5 provide a valid backlog for near-future research and development. The core issue of FP – the refugee crisis and global migration – will probably remain a challenge of high urgency. Global socio-political developments may yield new migrant waves soon, and the challenge of accommodating refugees in cities of the destination countries remains acute. Beyond the Hamburg case, MIT's Changing Places Group and HCU's CityScienceLab will promote the solution to other cities facing similar issues, e.g. European 'Arrival Cities' like Thessaloniki, Patras, Messina or Amadora. Alongside the FP effort, CS platforms could provide a large spectrum of applications to cities worldwide. CS capabilities span from urban planning, architecture and real estate development to logistic, data analysis and human-dynamics. Deploying CS platforms in Living Labs such as Hamburg CityScienceLab enables a fruitful exchange between academic research and real-life challenges of the hosting cities.

## References

[1] United Nations High Commissioner for Refugees (UNHCR) (2016): Global Trends. Forces Displacement in 2015
[2] Bundesamt für Migration und Flüchtlinge (BAMF) (2016): Das Bundesamt in Zahlen 2015. Asyl, Migration und Integration
[3] http://www.hamburg.de/pressearchiv-fhh/4505220/2015-06-02-pr-hafencity-universitaet/ (retrieved 05/24/2017)
[4] https://www.media.mit.edu/members/becoming-a-member-company/ (retrieved 05/24/2017)
[5] Freie und Hansestadt Hamburg/Zentraler Koordinierungsstab Flüchtlinge (2016): Schaffung von Unterkünften zur Flüchtlingsunterbringung durch die Freie und Hansestadt Hamburg. Monitoringbericht
[6] Ben Joseph, E.; Ishii, H.; Underkoffer, J.; Piper, B.; Yeung, L. (2001) Urban Simulation and the Luminous Planning Table: Bridging the Gap between the Digital and the Tangible, in: Journal of Planning Education and Research 21:196-203
[7] Larson, K. (2000) Louis I. Kahn – Unbuilt Ruins. Monacelli Press
[8] Noyman, A. (2015) Powerstructures: The form of urban regulations. Thesis MIT Cambridge
https://dspace.mit.edu/handle/1721.1/99301 (retrieved 05/24/2017)
[9] Ishi, H.; Ben-Joseph, E.; Underkoffer, J.; Yeung, L.; Chak, D.; Kanji, Z.; Piper, B. (2002) Augmented Urban Planning Workbench: Overlaying Drawings, Physical Models and Digital Simulation, in: Proceedings of IEEE & ACM ISMAR 2002, September 30 - October 1, 2002
[10] Hadhrawi, M., Larson, K. (2016) Illuminating LEGOs with Digital Information to Create Urban Data Observatory and Intervention Simulator, in: Proceedings of the 2016 ACM Conference Companion Publication on Designing Interactive Systems, 105-108
[11] Alrashed, T.; Almalki, A.; Aldawood, S.; Alhindi, T.; Winder, I.; Noyman, A.; Alfaris, A.; Alwabil, A. (2015) An Observational Study of Usability in Collaborative Tangible Interfaces for Complex Planning Systems, in: Procedia Manufacturing Volume 3, 2015, 1974-1980
[12] Alrashed, T., Almalki, A.; Aldawood, S.; Alfaris, A.; Al-Wabil, A. (2015) Coding Schemes for Observational Studies of Usability in Collaborative Tangible User Interfaces, HCI2015 International Conference on Human-Computer Interaction, Springer 2015 http://cces-kacst-mit.org/publication/coding-schemes-observational-studies-usability-collaborative-tangible-user-interfaces (retrieved 05/24/2017)